\RequirePackage[2020-02-02]{latexrelease}
\documentclass[times, twoside]{henriques-style}
\usepackage{blindtext}
\usepackage[nolist]{acronym}
\setcitestyle{square}
\usepackage{tikz}
\usetikzlibrary{positioning, calc, matrix, decorations.pathreplacing, shapes.geometric}
\usepackage{adjustbox}
\usepackage{times}
\usepackage{epsfig}
\usepackage{graphicx}
\usepackage{amsmath}
\usepackage{amssymb}
\usepackage{url}

\usepackage{booktabs}
\usepackage{multirow}

\usepackage{xspace}
\makeatletter
\DeclareRobustCommand\onedot{\futurelet\@let@token\@onedot}
\def\@onedot{\ifx\@let@token.\else.\null\fi\xspace}

\leadauthor{Shephard} 

\begin{document}

\title{TIAger: Tumor-Infiltrating Lymphocyte Scoring in Breast Cancer for the TiGER Challenge}
\shorttitle{TIAger Submission}

\author[*1]{Adam Shephard}
\author[*1]{Mostafa Jahanifar}
\author[1]{Ruoyu Wang}
\author[1]{Muhammad Dawood}
\author[1]{Simon Graham}
\author[2]{Kastytis Sidlauskas}
\author[3]{Syed Ali Khurram}
\author[1]{Nasir Rajpoot}
\author[1]{Shan E Ahmed Raza}
\affil[*] {Joint first authors contributed equally.}
\affil[1]{Tissue Image Analytics Centre, Department of Computer Science, University of Warwick, Coventry, UK}
\affil[2]{Barts Cancer Institute, Queen Mary University of London, London, UK}
\affil[3]{School of Clinical Dentistry, University of Sheffield, Sheffield, UK}
\maketitle

\begin{abstract}

The quantification of tumor-infiltrating lymphocytes (TILs) has been shown to be an independent predictor for prognosis of breast cancer patients. Typically, pathologists give an estimate of the proportion of the stromal region that contains TILs to obtain a TILs score. The Tumor InfiltratinG lymphocytes in breast cancER (TiGER) challenge, aims to assess the prognostic significance of computer-generated TILs scores for predicting survival as part of a Cox proportional hazards model. For this challenge, as the TIAger team, we have developed an algorithm to first segment tumor \textit{vs.} stroma, before localising the tumor bulk region for TILs detection. Finally, we use these outputs to generate a TILs score for each case. On preliminary testing, our approach achieved a tumor-stroma weighted Dice score of 0.791 and a FROC score of 0.572 for lymphocytic detection. For predicting survival, our model achieved a C-index of 0.719. These results achieved first place across the preliminary testing leaderboards of the TiGER challenge.
\end{abstract}



\begin{corrauthor}
adam.shephard@warwick.ac.uk, shan.raza@warwick.ac.uk
\end{corrauthor}

\setlength{\parindent}{15pt} 

\section*{Introduction}

Breast cancer is the most common type of cancer worldwide. In the UK, it accounts for 15\% of all new cancer cases and 7\% of all cancer deaths \cite{cruk2022}. Only 76\% of women diagnosed with beast cancer will survive their disease for ten years or more and it is still the second most common cause of cancer deaths in the UK \cite{cruk2022}. Breast cancer is primarily classified by its histological appearance, with most breast cancers deriving from the epithelial lining of the lobules or ducts (known as lobular/ductal carcinoma). 
It is a heterogenous disease, consisting of many types, and thus many factors must be considered when prescribing treatment for patients.

There has been a surge in literature emphasising the prognostic and predictive importance of tumor-infiltrating lymphocytes (TILs) visually assessed by pathologists on biopsies and surgical resections, particularly within triple negative (TNBC) and human epidermal growth factor receptor 2 (HER2+) breast cancer \cite{SALGADO2015, DENKERT2018}. Studies have shown that an increased degree of lymphocytic infiltration is prognostic of better long-term disease control. This has resulted in the development of guidance for generating a `TILs score'. Here, the proportion of stromal area (within the borders of the invasive tumor) to contain TILs is estimated and given as a TILs score between 0 and 100 \cite{SALGADO2015}.

With the advances of deep learning methods over the last decade for image analysis, several methods have been proposed to detect and/or segment nuclei and different regions in H\&E stained whole slide images (WSIs) \cite{Shephard2021, GRAHAM2019101563, jahanifar2022mitosis, aubreville2022mitosis}. 
The Tumor InfiltratinG lymphocytes in breast cancER (TiGER) challenge was formed to evaluate new computer algorithms for the automated assessment of TILs in HER2+ and TNBC histopathology slides. The challenge aims to inspire the generation of algorithms that can automatically generate a TILs score with high prognostic value, in an objective and efficient manner. As part of this challenge contestants were asked to submit an algorithm to two leaderbaords. Leaderboard 1 (L1) assesses the performance of the algorithm for the segmentation of tumor/stroma and the detection of all TILs in provided regions of interest (ROIs). Leaderboard 2 (L2) consists of WSIs that must be used to generate a TILs score for the case that is predictive of survival.

In this work, we propose a pipeline for use with breast cancer H\&E slides that segments tumor/stroma regions and detects TILs before producing a TILs score per slide. This algorithm was submitted to the TiGER challenge as team \textbf{TIAger}. Following the segmentation of tumor/stroma regions, our algorithm calculates the total tissue area from the tissue mask to determine whether the tissue should be part of the L1 or L2 pipeline. If the tissue is less than 5 mm\textsuperscript{2} in area, then it is processed in the L1 pipeline to segment tumor-stroma and detect all TILs in the entire tissue mask. If it is greater than this area, then it is processed as part of L2 that next generates a tumor bulk region and detects TILs within the stroma area of the tumor bulk region alone. These segmentations and detections are then used to generate a TILs score for the WSI. We first provide a description of the data used, before describing the segmentation/detection deep learning models and how they were trained. We next give a detailed overview of the whole pipeline for inference on new images along with the preliminary results achieved for L1 and L2. 
\newpage
We briefly list our contributions from this work below:

\begin{enumerate}
    \item First, we train segmentation and detection deep learning models based on the Efficient-UNet architecture to generate state-of-the-art results on the preliminary leaderboards.
    \item We then incorporate our models as part of a fully automated pipeline for TILs scoring.
    \item To allow our pipeline to be reproduced by future researchers we have made our code and model weights publicly available on GitHub: \url{https://github.com/adamshephard/TIAger}.
\end{enumerate}

\section*{Methodology}

\subsection{Data}
For training our segmentation and detection models, we employed the `WSIROIS' dataset. This dataset provided by the challenge organisers consists of 195 WSIs of breast cancer (core-needle biopsies and surgical resections) with pre-selected ROIs and manual annotations. This dataset was curated by combining cases from three sources:

\begin{enumerate}
\item \textbf{TCGA:} TNBC cases from TCGA-BRCA archive (n = 151). The annotations provided for this dataset were generated by adapting the publicly available BCSS \cite{bcss2019} and NuCLS \cite{nucls2021} datasets.
\item \textbf{RUMC:} 26 cases of TNBC and HER2+ cases from Radboud University Medical Center (Netherlands), with annotations by a panel of board-certified breast pathologists.
\item \textbf{JB:} 18 cases of TNBC and HER2+ cases from Jules Bordet Institute (Belgium). Annotations for these were made by a panel of board-certified breast pathologists.
\end{enumerate}

All data was extracted and provided at $20\times$ magnification (resolution of 0.5 micron-per-pixel) by the organisers for processing. All provided images have corresponding tissue masks for use in processing. Each image had manual region annotations provided with the following classes: invasive tumor, tumor-associated stroma, \textit{in-situ} tumor, healthy glands, necrosis not \textit{in-situ}, inflamed stroma and other. In this work, we merged the classes `tumor-associated stroma' and `inflamed stroma' to form a `stroma' class. We then set all other classes to zero, except the `\textit{in-situ} tumor' class. In this way, our segmentation task only consisted of three classes: background, stroma and tumor.

For the development and optimisation of our TILs score pipeline (e.g., generating tumor bulk region and forming a TILs score) we used the `WSITILS' dataset. This dataset was additionally provided by the challenge organisers and contained 82 WSIs of biopsies and surgical resections of TNBC and HER2+ breast cancer tissue from  RUMC and JB. Here, ground truth TILs scores were provided for each WSI from a board-certified breast pathologist.

\newpage
\subsection{Segmentation}
\subsubsection{Network Architecture}
\label{sec:architecture}
We employed a lightweight segmentation model, called Efficient-UNet \cite{jahanifar2021semantic}, for the segmentation task (implemented in TensorFlow). Efficient-UNet is a fully convolutional network based on an encoder-decoder design paradigm where the encoder branch is the B0 variant of Efficient-Net \cite{tan2019efficientnet}. Using this model with pre-trained weights from \textit{ImageNet} as a backbone allows the overall model to benefit from transfer-learning, by extracting better feature representations and gaining higher domain generalizability.
The following Jaccard loss function was ustilised to train the model \cite{jahanifar2018segmentation}:
\begin{equation}
{{\cal L}_{Jaccard}}\left( {{y_t},{y_p}} \right) = 1 - \frac{{\sum {{y_t}{y_p} + \epsilon } }}{{\sum {y_t^2 + \sum {y_p^2 - \sum {{y_t}{y_p} + \epsilon } } } }},
\label{eq:loss}
\end{equation}
where $y_p$, $y_t$, $\epsilon$ are the model prediction, ground truth segmentation, and a constant equal to 1  to avoid division by zero, respectively.

\subsubsection{Model Training}
\label{sec:training}
We trained the Efficient-UNet to semantically segment the input image into three prediction maps: invasive tumor (label index 1), stroma (label indices 2 and 6), and others (which comprises the background and all other tissue components rather than the first two).
The model was trained via a stratified 5-fold cross-validation framework, where each fold contained approximately the same number of images from each of the three subsets of the `WSIROIS' dataset. For training, patches of size $512\times512$ pixels were extracted at $10\times$ magnification, with a stride of 256 pixels. Images were normalised prior to being fed into the model for training by subtracting the ImageNet mean intensity and dividing by the ImageNet standard deviation. The segmentation model was trained using standard data augmentation techniques, with the addition of stain augmentation. The extent and combination of these augmentation techniques were randomly selected on-the-fly and differ from epoch to epoch.

Model training was performed across two phases. First, weights of the encoder were fixed to train the randomly initialised decoder for 10 epochs (Adam optimiser with learning rate of 0.003). We then trained the entire network for an additional 50 epochs using the same optimizer but a reduced learning rate of $4\times10^{-4}$. 

\subsubsection{Model Inference}

For model inference, we used the top three performing segmentation models from 5-fold cross-validation to obtain an ensembled result. Patches of size $512\times512$ were extracted from the tissue region at $10\times$ magnification with a stride of 256 and a zero-padding of 128 pixels. Inference was then performed on each patch using each of the three segmentation models. The predicted segmentation maps were then averaged and the resulting segmentation mask was then thresholded for each predicted tissue type ($\tau_{stroma} = 0.35$, $\tau_{tumor} = 0.20$,  optimised over experiments) and morphological opening was performed on the predicted tumor region (disc of radius 10). Following this, each segmentation mask was combined with the corresponding tissue mask. Only the centrally cropped $256\times256$ region was saved to the overall segmentation mask.

\subsection{Detection}
\subsubsection{Network Architecture}

For the detection task, we used the same model (Efficient-UNet) architecture and loss function as was employed for the segmentation task as described in Section \ref{sec:architecture}. The model takes an RGB input image of $128\times128$ pixels at $20\times$ magnification for inference and returns a $128\times128$ segmentation map.

\subsubsection{Model Training}

The model was trained via a stratified 5-fold cross-validation framework, where each fold contained approximately the same number of images from each of the three subsets of the `WSIROIS' dataset. A binary segmentation was produced for each TIL by dilating each ground truth detection by a radius of 5 pixels using morphological operations. In producing pseudo-segmentation masks for each detection we transformed the detection task into a segmentation task. For training, patches of size $128\times128$ were extracted at $20\times$ magnification, with a stride of 100 pixels. The same image normalisation,  data augmentation, loss function and two-phase training procedure were employed as in Section \ref{sec:training}.

The training dataset for the detection task had a severe class-imbalance, owing to much fewer extracted patches containing TILs than background. To alleviate this challenge, we devised an on-the-fly under-sampling approach which guarantees that the number of patches containing TILs and the ones that do not are almost equal in all batches.

\subsubsection{Model Inference}

For model inference we used the top three performing models from the 5-fold cross-validation and ensembled their output. Tiles of size $1024\times1024$ were extracted from the tissue region at $20\times$ with a stride of 1024 on inference. The patch was then further divided into sub-patches of size $128\times128$ with an overlap of 28 pixels. These patches were then fed into each of the three detection models. The output segmentation maps were then aggregated across each $1024\times1024$ tile. The output tile segmentations were then averaged to generate an ensembled lymphocyte segmentation map. We then thresholded the output segmentation mask (threshold = 0.3) and performed connected components analysis to find each individual detection. The centroids of the predicted nuclei and the mean intensities were then stored as the output detection coordinates/probabilities.

\subsection{Processing Pipeline}

The first step of our pipeline for processing each WSI consisted of performing semantic segmentation of the WSI. Next, we used the tissue mask to determine the area of the WSI. For L1, only ROIs were provided for processing, and we returned a semantic segmentation and a list of detections for the entire ROI area. In contrast, for L2, we were required to instead use the segmentations/detections to generate a TILs score for the WSI. Thus, if the area of the tissue mask was less than 5 mm\textsuperscript{2}, we performed the L1 pipeline and if it was greater, the L2 pipeline.

For the L2 pipeline, following the generation of the segmentation mask for the WSI we next found the `tumor bulk' region. This was defined using the same pipeline as the baseline algorithm (e.g., morphological operations followed by Delaunay triangulation). Following this, the tumor associated stroma was found by taking the overlap between the bulk tissue mask and the stroma from the inferred segmentation mask. We then performed detection in this stroma region alone. Next, we performed slide-level non-maxima suppression on the detected lymphocytes (patch size $2048\times2048$ pixels, at $20\times$ magnification). We then counted the number of lymphocytes within the bulk-stroma area. We multiplied this by an estimated lymphocyte area of 16 microns (optimised on `WSITILs') and calculated the area of the bulk stroma. The TILs score was calculated by dividing the total predicted TILs area by the stromal area and multiplying by 100. The score was further constrained to be an integer between 0 and 100. 

\section*{Results}

\begin{figure*}
\begin{center}
\includegraphics[width=17cm]{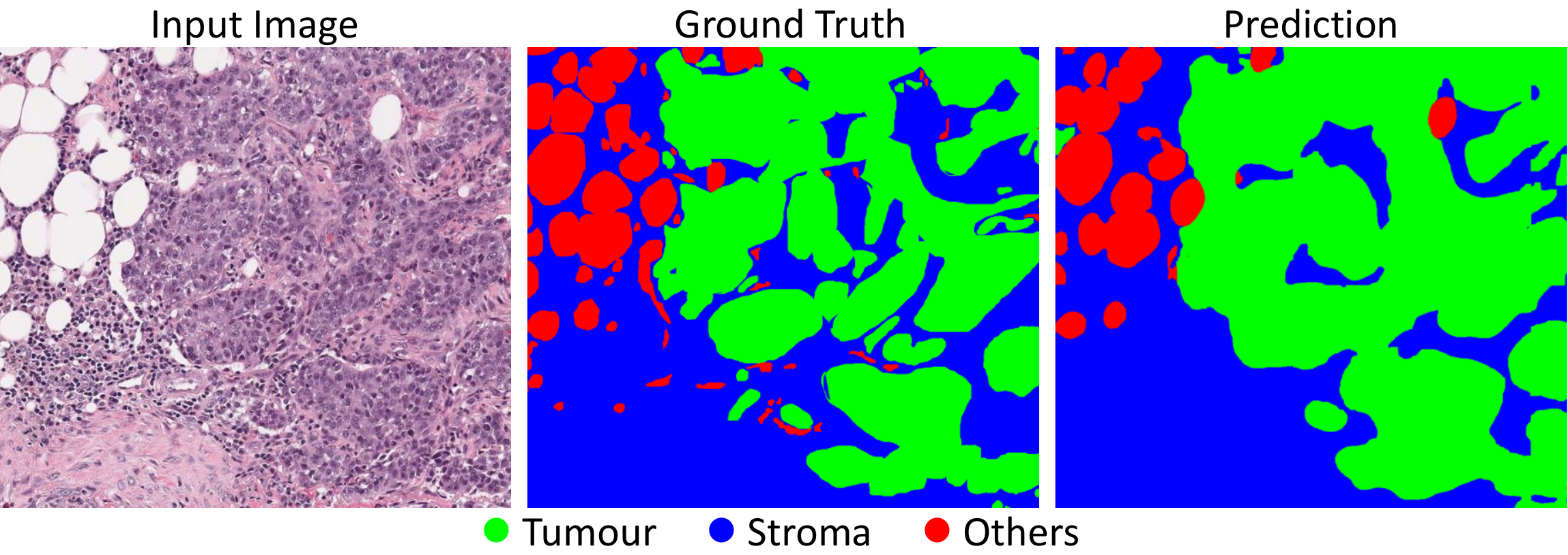}
\end{center}
   \caption{Sample segmentation output from Efficient-UNet with tumor, stroma and other regions coloured in green, blue and red, respectively.}
\label{fig:seg}
\end{figure*}

\begin{figure*}
\begin{center}
\includegraphics[width=17cm]{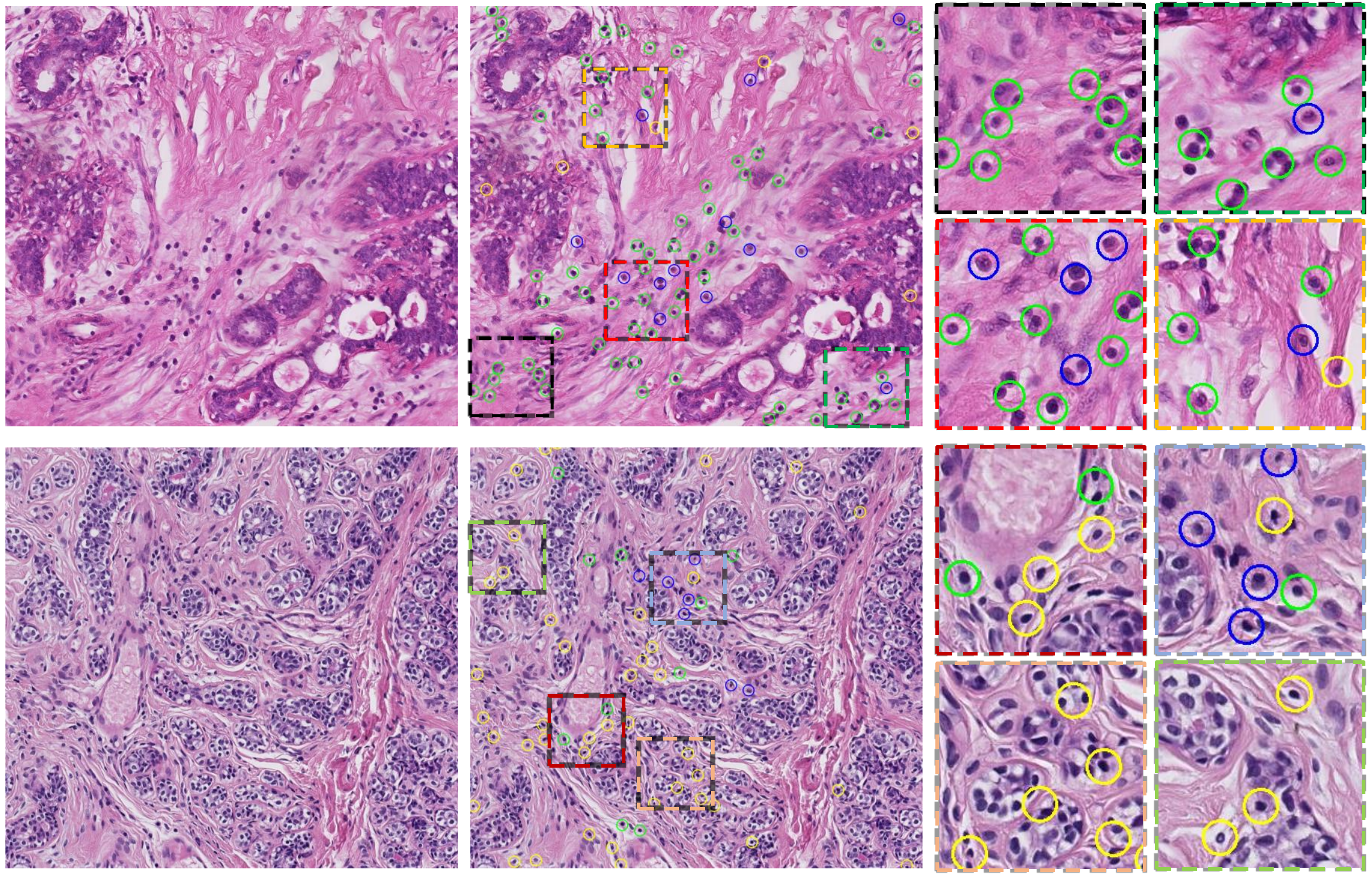}
\end{center}
   \caption{Sample detection output from Efficient-UNet: Raw input images (left); Detections (middle) shown in circles, where green circles represent true positives, blue circles false negatives and yellow circles false positives; Images showing further zoomed versions of the selected regions (right).}
\label{fig:det}
\end{figure*}

Many configurations for the segmentation and classification networks were tested, but the ones with minimum segmentation loss and best F1-score/dice scores across each experiment on 5-fold cross-validation were selected. The threshold values and hyper-parameters used for each detection/segmentation model were selected based on the cross-validation experiments. Following 5-fold cross-validation, our segmentation model achieved a dice score of 0.762 for tumor segmentation and 0.718 for stroma segmentation. We achieved a F1-score of 0.702 for detection. Example segmentation and detections with Efficient-UNet can be seen in Figures \ref{fig:seg} and \ref{fig:det} respectively. When predicting TILs scores with the `WSITILS' dataset we achieved a Pearson correlation coefficient of 0.744.

The preliminary test sets for L1 consisted of 26 WSIs with ROIs manually selected by the organisers. Using the proposed method, we achieved a tumor-stroma dice score of 0.791 on the L1 preliminary test set. Here, the tumor-stroma dice score is the average of the tumor vs background and the stroma vs background dice scores. For the detection of TILs we achieved a free-response receiver operating characteristic curve (FROC) score of 0.572 on the preliminary test set. When tested on L2 (containing 200 WSIs), our approach got the highest C-index of 0.719 for predicting survival as part of a Cox proportional hazards model based on our generated TILs scores. This demonstrates the robustness and generalisability of our method to new data. Combined, these results obtained first place in the challenge for both the L1 and L2 preliminary leaderboards.

\section*{Discussion and Conclusions}

In this work, we have presented a new pipeline for TILs scoring in breast cancer cases using the segmentation of tumor/stroma regions and the detection of TILs for each WSI. Both our segmentation and detection models are based on the Efficient-UNet architecture. The proposed method achieved the highest rank on the preliminary leaderboards of the TiGER challenge for both the segmentation/detection and predicting survival tasks.

\section*{References}

{\small
\bibliography{egbib}
}

\end{document}